\documentclass[12pt]{article}
\usepackage[left=2.54cm,top=2.54cm,right=2.54cm]{geometry}
\usepackage{fontenc}
\usepackage{graphicx}
\usepackage{float}
\usepackage{amssymb, amsthm}
\usepackage{hyperref}
\usepackage{fancyhdr}
\usepackage{listings}
\newtheorem{prop}{Proposition}
\begin{document}
\title{Oligopoly Dynamics}

\author{Bernardo Melo Pimentel\thanks{ISEG, U. of Lisbon\href{mailto:bmpimentel@iseg.ulisboa.pt}{\tt{<bmpimentel@iseg.ulisboa.pt>}}.
}}

\date{Fall 2017}

\maketitle
\begin{abstract}
  The present notes summarise the oligopoly dynamics lectures professor Lu\'is Cabral gave at the Bank of Portugal in September and October 2017. The lectures discuss a set industrial organisation problems in a dynamic environment, namely learning by doing, switching costs, price wars, networks and platforms, and ladder models of innovation. Methodologically, the materials  cover analytical solutions of known points (e.g., $\delta =0$), the discussion of firms' strategies based on intuitions derived directly from their value functions with no model solving, and the combination of analytical and numerical procedures to reach model solutions. State space analysis is done for both continuous and discrete cases. All errors are my own.
\end{abstract}

\section{Learning by doing}
\subsection{Motivation}
Learning by doing, or learning economies, was first documented in the aviation industry. In 1965, Boeing and PanAm agreed on the development of a new, wide body aircraft, the B747. Days later, F. Kolk from American Airlines sends manufacturers his own specifications. These were for a smaller plane than the 747. Lockheed, a military contractor and the inventors of most supersonic technology decided to enter the race. So does McDonnell Douglas and first proposals were submitted by September 1967. All three planes were similar, leading to fierce competition for launch orders. The average difference between orders during the initial period was estimated to be in the \$100,000 value for prices that fell in the \$15-17m range, less than 20\% above their cost. Basically, the situation was that of a Bertrand competition model.

Benefiting from being first delivering the design, Boeing took the initial lead. The main rival, Lockheed, made a splash with its technologically sophisticated design and thus captures a significant share of the market for itself. Pilots in particular, were keen about the Lockheed L1011. The Lockheed plane used British Rolls-Royce engines and as this supplier ran into trouble, Lockheed began experiencing long delays in its L1011 deliveries. This led to order cancellations and increase in the market share of Boeing. With a greater number of orders, Boeing benefit from a learning curve effect, making itself more competitive which lead to a self-reinforcing dynamic. Lockheed attempted a comeback but it was too late.

It is a well-documented fact that aircraft production costs significantly decline with cumulative production ($\eta \approx -0.3$). Under these conditions, an early lead may lead to a sustained lead with hyper-competition in initial stages and market power at later stages. That was the case of the wide body aircraft industry from the late 1960s to the 1990s. By the mid-1980s both Lockheed and McDonnell Douglas had exited the market. Only Boeing and the EU-backed entrant Airbus remained in operation in the wide body aircraft market.

\subsection{General model}
The assumptions for the learning by doing model are as follows:

\begin{itemize}
  \item infinite discrete periods, discount factor $\delta$;
  \item two sellers, one consumer per period;
  \item state ($i,j$: cumulative sales by firms $i$ and $j$);
  \item timing (within each period):
  \begin{itemize}
    \item firms simultaneously set prices $p(i,j)$;
    \item consumer makes purchase
    \item state is updated according to sales
  \end{itemize}
  \item equilibrium concept: Markov perfect equilibrium (MPE).
\end{itemize}

The learning by doing effect is modelled under the assumptions
\renewcommand{\labelenumi}{(\roman{enumi})}
\begin{enumerate}
  \item $c(i+1) < c(i)$  for $m>i>0$;
  \item $c(i) = c(m)$ for $i>m$.
\end{enumerate}

The consumer buys a single unit per period with no outside option. The sale generates $\zeta_i$ value for firm $i$ so that $\xi_i = \zeta_i-\zeta_j$ is the difference in value firm $i$ gets in relation to firm $j$ from selling . $F(\xi)$ is the c.d.f. of $\xi_i$ under the following assumptions

\renewcommand{\labelenumi}{(\roman{enumi})}
\begin{enumerate}
  \item $F(\xi)$ is the c.d.f. and is continuously differentiable;
  \item the p.d.f. is symmetric so that $f(\xi) = f(-\xi)$;
  \item $f(\xi) >0, \forall \xi $;
  \item $F(\xi)/f(\xi)$ is strictly increasing.
\end{enumerate}

Furthermore, the consumer chooses firm $i$ iff $\zeta_i-p(i,j)\geq\zeta_j-p(j,i)$. This happens with probability $q(i,j)$, so that the demand faced by firm $i$ is $q(i,j) = 1 - F(P(i,j))$. As the price of $i$ increases, its demand $q(i,j)$ decreases. We will define the price difference between $i$ and $j$ as being equivalent to

\begin{equation}\label{pricediff}
P(i,j)\equiv p(i,j) - p(j,i)
\end{equation}

Also
\begin{itemize}
  \item $P(j,i) = - P(i,j)$;
  \item $q(j,i) = F(P(i,j))$;
  \item $d q(i,j)/d p(i,j) = -f(P(i,j))$.
\end{itemize}

Writing down the Bellman equation for firm $i$'s value function and the respective first-order conditions we have

\begin{equation}\label{bellman_v1}
  v(i,j) = \Big( 1-F\big(P(i,j)\big)\Big) \Big(p(i,j) - c(i) + \delta v(i+1,j) \Big) + F\Big( P(i,j)\Big) \delta v(i,j+1)
\end{equation}

where $\Big( 1-F\big(P(i,j)\big)\Big)$ is $i$'s demand if it sells, $p(i,j)$ is the price, $c(i)$ the unit cost, and $\delta v(i+1,j)$ the discounted increased value from selling. If firm $i$ is unable to sell, then it faces no demand in the period $F\Big( P(i,j)\Big)$ and firm $j$ is the one increasing in future value $\delta v(i,j+1)$.

Taking the first order condition we have that

\begin{equation}
  0 = 1 - F\big(P(i,j)\big) - f \Big( P(i,j) \Big) \Big( p(i,j) - c(i) + \delta v\big(i+1,j \big) - \delta v\big(i,j+1\big) \Big).
\end{equation}

Solving for $p(i,j)$, we get the equilibrium price

\begin{equation}\label{zeroprofit}
  \hat{p}(i,j) = c(i) + \frac{1-F\big( P(i,j) \big)}{f\big(P(i,j) \big)} - \delta w(i,j)
\end{equation}

where $c(i)$ is the unit cost and

\begin{equation}
\frac{1}{\eta} =  \frac{1-F\big( P(i,j) \big)}{f\big(P(i,j) \big)}
 \end{equation}

is the inverse price elasticity of demand. The higher the elasticity, the lower the price charged by firm $i$.

\begin{equation}\label{vdiff}
w(i,j) \equiv v(i+1,j) - v(i,j+1)
 \end{equation}

is a dynamic component representing the profit premium firm $i$ earns over $j$. It is hence the difference in between the value of $i$ and $j$. There is an inter-temporal subsidy component in the last components of (\ref{zeroprofit}), as firm $i$ is willing to charge lower prices today in order to guarantee greater future earnings. The greater the $\delta w(i,j)$ terms, the smaller the price and the more significant the current price subsidy for future earnings is.

The motion equations linking the prices of firms $i$ and $j$ then are

\begin{eqnarray*}
\hat{p}(i,j) &=& c(i) + \frac{1-F\big( P(i,j) \big)}{f\big(P(i,j) \big)} - \delta w(i,j) \\
\hat{p}(j,i) &=& c(j) + \frac{1-F\big( P(j,i) \big)}{f\big(P(j,i) \big)} - \delta w(j,i)
\end{eqnarray*}

from (\ref{pricediff}) we get that

\begin{equation}\label{motioneq}
  \hat{P}(i,j) = c(i)-c(j) + \frac{1-2F\big(P(i,j) \big)}{f\big(P(i,j) \big)} - \delta \Big(w(i,j) - w(j,i) \Big)
\end{equation}

where $c(i)-c(j)$ is cost difference and $\Big(w(i,j) - w(j,i) \Big) $ is the difference-in-difference in profits of the two firms \footnote{from equation (\ref{vdiff})}. Let then

\begin{equation}
  K'(P(i,j)) = C(i,j) - \delta W(i,j)
\end{equation}

where $C(i,j)\equiv c(i) - c(j)$ and $W(i,j) \equiv w(i,j) - w(j,i)$. $K'(x)$ is monotonic such that $K'(x)>0$ and $K(0) = 0$.

Modelling the pricing dynamics, suppose, without loss of generality, that $i\geq j$. Then the probability that the next sale goes to the leader $i$, thus further increasing its dominance is $q(i,j)>q(j,i)$ iff

\begin{equation}\label{increasedom}
  C(i,j) - \delta W(i,j) < 0.
\end{equation}

Equation (\ref{increasedom}) shows that two factors are at play when determining the dominance of firm $i$. First, a cost advantage over $j$ is secured if the first component is such that $C(i,j)\leq 0$. Second, price cutting incentives exist such that $W(i,j) \lesseqgtr 0$. The greater the desire for form $i$ to win the next sale, the larger this term will be. The combination of the two terms gives an interpretation that is analogous to the classic mechanics principle of least action.

We conclude the analysis of the general model by going back to the value function, now in equilibrium

\begin{equation}
  \hat{v}(i,j) = H\big(P(i,j) \big) + \delta v (i,j+1)
\end{equation}

where

\begin{equation}\label{staticpi}
H(x) \equiv \frac{\Big(1 - F(x) \Big)^2}{f(x)}.
\end{equation}

Firm $i$'s equilibrium value is thus dependent on todays prices -- via the $H\big(P(i,j) \big)$ term -- and tomorrow's hyper-competitive outlook -- given by $j+1$. Equation (\ref{staticpi}) corresponds to the ``static'' equilibrium profit. The continuation value equals that from losing the current sale. This is a Bertrand style of competition: any extra gain is competed away in a price war. As a final methodological point, note that the $\hat{v}(i,j)$ value function can be solved sequentially.

\subsection{Increasing dominance and hyper-competition}
We now explore an increasing dominance by a market leader using a two-step learning case. If firm $i$, the current leader, has experience its marginal cost is $c(i) = c(1), \forall i\geq 1$. As before, only another firm exists, the laggard. This yield four possible scenarios, under which both, none, the leader, or the laggard benefit from experience

\begin{eqnarray*}
    v(1,1) &=& \frac{H(0)}{(1-\delta)} \\
    v(0,1) &=& \frac{H\big(P(0,1)\big)}{(1-\delta)} \\
    v(1,0) &=& \frac{\delta H(0) + (1-\delta) H\big( P(1,0) \big) }{(1-\delta)} \\
    v(0,0) &=& \frac{(1-\delta)H(0) - \delta H\big( P(0,1) \big)}{(1-\delta)}
\end{eqnarray*}

since $P(j,i) = - P(i,j)$, we have only one unknown. Also, the master motion equation implies an unique solution for the system. Under these conditions, an increase in dominance will occur because the market leader is more likely to make the sale than the laggard. That is,

\begin{equation}
    P(1,0) = p(1,0) - p(0,1) <0.
\end{equation}

To proof the previous proposition we turn to the difference-in-difference in the profits between the leader and laggard firms. This is given by the difference in equilibrium profits from only the leader and the leader and the laggard having experience

\begin{equation}
    W(1,0) = H(1,0) - H(1,1) = \frac{\Big(1-F\big(P(1,0)\big)\Big)^2 }{f\big(P(1,0)\big)} - \frac{\big(1-F(0)\big)^2}{f(0)}.
\end{equation}

Hence

\begin{equation}\label{lhs}
    P(1,0) + \frac{2F\big(P(1,0)\big) - 1}{f\big(P(i,j)\big)} +  \frac{\Big( 1 - F\big(P(1,0)\big)\Big)^2}{f\big(P(1,0)\big)} - \frac{\big( 1- F(0)\big)^2}{f(0)} = c(1) - c(0)
\end{equation}

where the L.H.S. of (\ref{lhs}) is strictly increasing in $P(1,0) = 0$ when $P(1,0) =0$ and the R.H.S. is strictly negative, that is, $P(1,0)<0$. Hence, a unique solution exists and confirms the conjecture that the experienced firm faces lower costs than those of the laggard. The intuition is that the cost difference clearly benefits the leader. However, we still need to analyse the profit implications in $W$. The profit D.I.D. for the leader is

\begin{eqnarray}
    W(1,0) &=& w(1,0) - w(0,1) \\
           &=& \big( v(1,0) - v(1,1) \big) - \big( v(1,1) - v(0,1)\big) \label{profitcomp}
\end{eqnarray}

which contemplates the scenario under which the leader wins the next sale (the first term of \ref{profitcomp}) and the case in which the previous laggard now wins the sale (the second term of the equation). Hence, leader dominance  increases in  $W(1,0)>0$ iff

\begin{equation}
    v(1,0) + v(0,1) > v(1,1) + v(1,1)
\end{equation}

that is, only if the profits of one firm being in state 1 are greater than the profits of the two being in state 1 (being experienced).

In order to explore the hyper-competition scenario, we analyse the case of a steeper learning curve, so that $c(1)$ is lower. Let us fix $c(0)$ and lower the value of $c(1)$. What happens to the equilibrium pay-offs? Results will show that if costs decrease, so will firm value, with $v(0,0)$ being strictly increasing in $c(1)$. This means the value of both firms will be lower should none benefited from a learning effect. Formalising the proof

\begin{equation}
    P(1,0) = \frac{2F\big(P(1,0)\big)-1}{f\big(P(i,j)\big)} + \frac{\Big(1-F\big(P(1,0)\big)\Big)}{f\big(P(1,0)\big)} - \frac{\big(1-F(0)\big)^2}{f(0)} = c(1) - c(0)
\end{equation}

implies that $P(1,0)$ is increasing in c(1). When the firms have no experience we have

\begin{equation}
    v(0,0) = \frac{(1-\delta)H(0) - \delta H\big(P(0,1)\big)}{(1-\delta)}
\end{equation}

which implies that $v(0,0)$ is increasing in $P(1,0) = -P (0,1)$.

This hyper-competitive result may be better understood with the example of a first price auction in which the winner gets $\pi$ and the loser pays $-\pi$. Both bidders are forced to bid.  The equilibrium bid in the auction will be $2\pi$ ($\pi-2\pi$ for the winner and $-\pi-0$ for the loser). In the present context, the equilibrium continuation value is value conditional on losing the sale. A steeper learning curve implies a higher pay-off for leader and lower pay-off for the laggard. This is the classical Bertrand trap: the strategic effect exactly cancels the direct effect. With an even steeper learning curve we face a Bertrand supertrap: the strategic effect outweighs the direct effect, leaving the firms worse off should when entering the auction. In this case, the benefit from a cost reduction (the direct effect) is more than cancelled by the strategic effect that directs incentives both firms to lower their prices in an attempt to capture a larger share of the market.

Recent research attempts shed some light on the effect of corporate forgetting, that is, an increase in unitary costs from not producing a good. Empirical evidence suggests forgetting is an important phenomenon. However, the mathematical modelling of a forgetting effect has lead to multiple outcomes that are harder to make sense of. Analytical results are difficult to obtain, so numerical methods, especially homotopy have been attempted to moderate success.

\subsection{Predatory pricing}
We now move to the dynamics of predatory pricing. A firm may reduce its prices in an attempt to destroy its rivals or to deter new entry. The model's assumptions are

\begin{itemize}
    \item In each period nature determines a firm's avoidable fixed cost, either $0$ ($\Pr \alpha $) or i$A$ ($\Pr 1-\alpha$), where $A$ is a very large fixed cost;
    \item firms simultaneously decide wether to remain active or exit, with exit being irreversible;
    \item if both firms are active, then they simultaneously set price;
    \item let
        \begin{itemize}
            \item $\tilde{x}$ be the equilibrium value of $x$ in fixed cost case;
            \item $\hat{x}$ be the equilibrium value of $x$ in no fixed cost (no exit) case;
            \item  $\tilde{v}(i)$ be the monopoly value function.
        \end{itemize}
\end{itemize}

Using the previous assumptions, suppose that $\alpha$ is infinitesimally small, $\tilde{v}(1) \gg v(1,1)$ and $A = v(0,1)$. This is the case in which the fixed cost is virtually impossible to avoid, so the value of a single firm operating in monopoly ($\tilde{v}(1)$) is much greater than the value of a duopoly. In this case, an equilibrium with exit at $(0,1)$ exists if the fixed cost is positive. Moreover, large price differences exist, with the leader charging a significantly lower price $\tilde{P}(1,0)<P(1,0)$. In the limit, as $\alpha \rightarrow 0$

\begin{eqnarray*}
    \tilde{v}(1,1) &\rightarrow& v(1,1) \\
    \tilde{v}(0,1) &\rightarrow& 0\\
    \tilde{v}(1,0) &\rightarrow& \tilde{v}(1)
\end{eqnarray*}

Therefore, in the limit as $\alpha \rightarrow 0$

\begin{eqnarray*}
    \tilde{W}(1,0) &\rightarrow& \big(\tilde{v}(1) - v(1,1)\big) - \big(\tilde{v}(1,1) -0\big) \\
    \tilde{W}(1,0) &=& \tilde{v}(1) - 2v(1,1)
\end{eqnarray*}
which is positive and independent of $\tilde{P}(1,0)$. Generically, the price difference $P(1,0)$ is given by

\begin{equation}
    K\big(P(i,j)\big) = C(i,j) - \delta W(i,j)
\end{equation}

where $K(x)$ is strictly increasing and $K(0) = 0$. If no exit takes place, $\tilde{W}(i,j)$ is increasing in $\hat{P}(1,0)$ and zero if $\hat{P}(1,0) = 0$. In the case of exit $\tilde{W}(i,j) >0$ and independent of $\tilde{P}(1,0)$. It therefore follows that $\tilde{P}(1,0) < \hat{P}(1,0)$.

In equilibrium rational entry occurs if a positive value for the monopolist exists, that is, if $\tilde{v}(0,0)>0$. Predatory behaviour occurs if $\tilde{P}(1,0)<\hat{P}(1,0)$ (the Ordover-Willig model result). This means that price aggressiveness is higher because exit is a possibility. Exit is rational for the prey of predatory pricing, as $\tilde{v}(0,1) =0$ with $\Pr 1-\alpha$.

We find an equilibrium with entry, predation, and exit by rational players without a need to resort to asymmetric information assumptions. In equilibrium, the $p<MC$ is neither a necessary nor a sufficient condition for predation with multiple equilibrium being possible and, in fact likely. For this reason, the overall welfare effects of predation are ambiguous.

A number of views on predation have been put forward over the years. The Chicago school view is one of complete information in which no subgame perfect equilibrium with predation exists. A second take, by Migrom and Roberts \cite{milgrom1982predation}, considers the existence of bootstrap equilibria. In equilibrium, the prey exits because it expects the predator to continue to set a low $p$ in the future and the predator sets a low $p$ in the expectation that the prey will exit.

Under asymmetric information, two explanations for predatory pricing may be considered. The first is to create a reputation for toughness, as argued by Kreps and Wilson \cite{kreps1982reputation}, under which the predator sets a low price in order to drive out current rivals and preempt potential entry by other firms. The second is the deep purse argument of Bolton and Scharfstein \cite{bolton1990theory}, by which a predator sets a low price in order with the prospects of depleting the prey's profits and cash reserves. In this scenario, the prey will have a constrained access to financing, this exiting the market.

Finally, a number of predation models take advantage of dynamic profit functions to model scenarios of learning by doing, switching costs, and network effects. With $q_{it}$ influencing profits $\pi_{j,t+1}$ and with uncertainty concerning the future profit levels, aggressive pricing at time $t$ increases the probability of rival exit at time $t+1$. Examples of such models include Cabral and Riordan \cite{cabral1994learning} and Besanko, Doraszelski, and Kryukov \cite{besanko2013sacrifice}.

Summarising, we saw how learning curves lead to three important effects. First, they increase dominance of the leader over the laggard by price  convergence or divergence. Second, learning may induce hyper-competition as reductions in firms' costs lead to decreased firm value. Third, they also stimulate predatory pricing practices, under which lower prices by one firm push down the learning curve and increase chances of higher future market shares (or monopoly). These effects are of more general applicability: profit functions with inter-temporal dependencies.

\section{Switching Costs}
\subsection{Motivation}
Switching costs refer to impediments customers face when changing suppliers. Switching costs arise when exit fees, search costs, learning costs, cognitive effort, emotional costs, equipment costs, installation and start-up costs, financial risk, psychological risk, and social risk exist. Switching costs will vary with the intensity of rivalry in the industry. Economic theory and empirical results show that large switching costs reflect the market power of incumbents. However, in customer markets where the number of customers is relatively small and the seller has considerable information about the buyer, switching costs may induce competition rather than hamper it. Examples of such b2b markets include the ready-mixed concrete, the enterprise software and the tug boat push service industries. In such markets, although there is a list price (a rack rate), each customer receives a idiosyncratic discount. The final price then depends on the customer's ability to pay and bargaining power.

\subsection{General Model}
The model assumptions are
\begin{itemize}
	\item infinite periods, two sellers, one buyer
	\begin{itemize}
		\item sellers discriminate between locked-in and non locked-in buyers;
		\item this implies that there will exist one single buyer for each price;
	\end{itemize}
	\item timming within each period
	\begin{itemize}
		\item sellers simultaneously set prices;
		\item nature generates buyers i.i.d. preference shocks (utility) $\zeta_i (i=1,2)$;
		\item buyers purchases one unit from one of the sellers and pays switching cost $s$ if the seller is different from previous period seller;
	\end{itemize}
	\item symmetric Markov equilibria, where state is
	\begin{itemize}
		\item the buyer's observed preferences;
		\item identity of "incumbent", lock-in seller (the seller previously chosen by the buyer).
	\end{itemize}
\end{itemize}

The buyer's preferences are such that the outside option is not feasible (worth $-\infty$) and thus it buys one unit from one seller. There is a relative preference for firm $i$ at time $t$: $\xi_{it}$, where the relative preference for one seller is $\xi = \xi_i - \xi_j$ $i = A,B$; $t \in N$. $\xi_{it} \sim F(\xi_{it})$ i.i.d., buyer's private information is $\xi_{jt} = - xi_{it}, i \neq j$.

The cumulative distribution of the buyer's preferences $\xi_{it}$ is
\begin{itemize}
	\item continuously differentiable;
	\item symmetric around zero: $f(x) = f(-x)$;
	\item such that it has a positive density: $f(x)>0, \forall x$;
	\item unimodal: $f(x)$ has one mode (at zero);
	\item such that it has a monotone hazard rate (MHR): $F(x)/f(x)$ is strictly increasing.
\end{itemize}

Final assumption comes from the fact that the MHR implies the following to be strictly increasing in $x$

\begin{equation}
	\frac{F(x)^2}{f(x)},\quad \frac{F(x)-1}{f(x)},\quad \frac{2F(x)-1}{f(x)}.
\end{equation}

Moreover, the following is increasing in $x$ iff $x>0$ and constant at $x=0$

\begin{equation}
	\frac{\big(1-F(x)\big)^2 +\big(F(x)\big)^2 }{f(x)}.
\end{equation}

In the model, preferences are serially uncorrelated. That means that past preferences, {\it not purchases}, are temporally independent. An iPhone owner may prefer an Android phone today, even if she doesn't make a purchase of a phone today. We will also be dropping the $t$ subscript and replacing the $i =A,B$ with $i=0,1$ to denote "outsider" or "insider" state. Given symmetry, customers do not care about continuation value. Hence, a buyer will choose the "insider" iff its unobservable preferences for firm 1 are  $\xi_1 - p(1) \geq p(0) - s$, where $s$ is the switching cost. Let us also define $x \equiv p(1) - p(0) - s$ as the insider's relative price adjusted by switching costs. Then, the demand faced by the outsider firm $0$ is $q(0) = F(x),$. Demand for the insider is then $\quad q(1)=1-F(x)$.

The insider seller optimization process occurs along its value function

\begin{equation}
	v(1) = \big(1-f(x)\big) \big(p(1) + \delta v(1) \big) + F(x) \delta v(0).
\end{equation}

Its F.O.C. is

\begin{equation}
	-f(x) \big(p(1) +\delta v(1) \big) + \big(1 - F(x) \big) + f(x) \delta v(0) = 0
\end{equation}

and the optimal prices for both firms are

\begin{eqnarray}
	p(1) = \frac{1-F(x)}{f(x)} - \delta \big(v(1)-v(0)\big) \label{pricein}\\
	p(0) = \frac{F(x)}{f(x)} - \delta \big(v(1)-v(0)\big)  \label{priceout}
\end{eqnarray}

where $v(1)-v(0)$ is the investment effect from switching. The results from (\ref{pricein}) and (\ref{priceout}) are equivalent to the well-known elasticity rule\footnote{that is, $\epsilon = -\frac{\partial Q}{\partial P} \frac{P}{Q}$} $\epsilon_1 \equiv -f(x) \frac{p(1)}{1-F(x)}$. Two effects are therefore present: a harvesting and an investment effect. The switching costs increase the rigidity of demand and that leads to a desire for greater unit margins by the sellers. On the other hand, in order to create switching costs, firms need to increase the $v(1)-v(0)$ component so that the customers do not invest in the alternative offer in the future.

In equilibrium, we get the following value functions from substitution in the F.O.C.s

\begin{eqnarray}
    \hat{v}(1) = \frac{\big( 1-F(x)\big)^2}{f(x)} + \delta v(0) \\
    \hat{v}(0) = \frac{F(x)^2}{f(x)} + \delta v(0).
\end{eqnarray}

This is a dynamic version of the "Bertrand trap": the gain from winning a sale is bid away, so the continuation value is $v(0)$.

Solving the rest of the model from the F.O.C.s

\begin{eqnarray}
    p(1) &=& \frac{1-F(x)}{f(x)} - \delta \big(v(1) - v(0) \big) \\
    p(0) &=& \frac{F(x)}{f(x)} - \delta \big(v(1) - v(0) \big).
\end{eqnarray}

We defined $x\equiv p(1) - p(0) - s$. Taking the differences from the F.O.C.s

\begin{equation}
    x = \frac{1-2F(x)}{f(x)} - s
\end{equation}

which, given increasing $F(x)/f(x)$ implies an unique solution for $x$. Moreover, $x$ is decreasing in $s$ and so $q(1)$ is increasing in $s$. Intuitively, the probability of an insider selling on the current period positively depends on the switching cost $s$.

\subsection{Numerical model simulation}
Suppose the preference shocjs follow $F(x)$ a  standardized normal distribution. With $\frac{2F(x) - 1}{f(x)}$ being strictly increasing, we can use the intermediate value theorem to find the unique numerical solution $x$ to $LHS(x) =0$ where

\begin{equation}
    LHS = x + \frac{2F(x) - 1}{f(x)} + s.
\end{equation}

The {\tt Julia} script, with the {\tt Distributions} extension, is then

\begin{lstlisting}[language=python, frame=single, basicstyle=\small]
using Distributions
F(x) = cdf(Normal(0,1),1000)     # Normal cdf
f(x) = pdf(Normal(0,1),1000)     # Normal pdf
xL = -1e3                        # Initial bounds of x
xH = +1e3
x = (xL + xH)/2                  # Try middle value of x
s = 5                            # Value for s
LHS = x + (2 + F(x) - 1)/f(x) + s
while (abs(LHS) > 1e-5)          # Main loop; precision = 1x10^(-5)
    if (LHS > 0)                 # If LHS>0, root must be to the left
        xH = x
    else
        xL = x
    end
    println(x)
end
\end{lstlisting}

\subsection{Effect of $s$ on competition}
Higher switching costs lead to higher prices for the insider so that $p(1)>p(0)$. If we define the average price as

\begin{equation}
    \bar{p} \equiv p(0) q(0) + p(1)q(1)
\end{equation}

where demand $q(0,1)$ is a probability. Then

\begin{prop}
    There exist values $s',s''$, where $0<s'<s''<\infty$ so that \\
    (a) $s<s'$, then average price $\bar{p}$ is decreasing in switching cost $s$ \\
    (b) $s>s''$, the average price $\bar{p}$ is increasing in switching cost $s$.
\end{prop}

To prove the proposition we start by taking the differences from the value functions

\begin{equation}
    V \equiv v(1) - v(0) = \frac{1-2F(x)}{f(x)}.
\end{equation}

Substituting this and plugging the F.O.C. into definition of $\bar{p}$

\begin{eqnarray*}
    \bar{p} &=& \big(1 - F(x)\big) \Big(\frac{1-F(x)}{f(x)} + \delta V \Big) + F(x) \Big(\frac{F(x)}{f(x)} - \delta V\Big) \\
             &=& \frac{\big(1-F(x)\big)^2 + F(x)^2}{f(x)}  + \delta \Big(\frac{2F(x)-1}{f(x)} \Big).
\end{eqnarray*}

The lemma implies that, at $x=0$ the first term on the right hand-side is constant in $x$, and the second increasing in $x$. Hence, if $x$ is small, then $d\bar{p}/d x >0$; and $d x/d s <0$.

The in intuition is as follows: for small switching costs $s$, consumers will be fairly indifferent between making the switch. Specifically, for a  small $s$, $\delta=0, s\rightarrow 0, q(0) = q(1) = \frac{1}{2}$. That is, firms evenly split the market. Moreover, $\partial p(1)/\partial s = - \partial p(0) /\partial s$, that is, as the switching cost increases the insider price increases and the outsider price decreases. For this reason, the average price $\bar{p}$ remains approximately constant. The value difference $v(1) - v(0)$ is increasing in $s$, so both prices $p(1)$ and $p(0)$ decrease with $s$.

The effects for when the switching cost is large are as expected. When $s$ is sufficiently large demand for the outsider is basically inexistent $q(0) \approx 0$ and the insider dominates the market $q(1) \approx 1$. Moreover, insider prices increase with the switching cost so that $ \partial p(1)/\partial s >0$. In this case, the average price $\bar{p}$ increases. The corollary for these observations is that for small switching costs, increases in s make the market more competitive, as prices decrease.

The  takeaways are that switching costs lead to a  bargain-then-ripoff price pattern. Also, switching costs magnify existing market conditions: they increase competitiveness of a already competitive markets and decrease competitiveness in little competitive markets. A meta game whereby firms chose their own switching cost may have the nature of a prisoner's dilemma. Finally, switching costs may increase consumer welfare but tend to decrease social welfare.
\nocite{*}
\bibliographystyle{siam}
\bibliography{oligopoly}

\end{document}